# GROUPS AND CLUSTERS IN THE NEAR-INFRARED

GARY A. MAMON

*DAEC, Observatoire de Paris-Meudon, F–92195 Meudon, FRANCE*

**Abstract.**
Massive multicolor digital surveys in the near infrared provide new opportunities for statistical analyses of the properties of groups and clusters. These galaxy systems provide a natural laboratory for studying dynamical processes and galaxy evolution, and serve as powerful cosmological diagnostics. The immediate scientific returns to be expected from near-infrared surveys such as DENIS and 2MASS are presented, and illustrated with simulated images of a Coma-like rich cluster.

## 1. Astrophysical Motivations

Thanks to gravity, galaxies like to congregate in groups and clusters. Although rare, clusters are popular as they are the largest objects whose cores are in dynamical (*virial*) equilibrium, as contamination by interlopers is not too significant. Loose groups of 4 to say 20 galaxies have the advantage of being numerous, and for this reason, are often used as distance indicators (gaining $N_{\rm gal}^{-1/2}$ in accuracy relative to individual galaxies). Compact groups of usually 4 galaxies appear so dense in projection that they would be the highest density isolated systems of galaxies, denser than the cores of rich clusters. Unfortunately they are very rare and may suffer from serious contamination from a surrounding loose group (Mamon 1986) or cluster (Walke & Mamon 1989).

## 2. Recent Developments on Clusters

Recent observations with the *ROSAT* X-ray satellite indicate that clusters are not as smooth as generally assumed earlier (*e.g.*, White, Briel & Henry 1993), while recent analyses of optical data point out to a similarly high frequency of small-scale substructure (Salvador-Solé, Samromà & Gonzalez-Casado 1993; Salvador-Solé, Gonzalez-Casado & Solanes 1993). Now, in a low $\Omega$ Universe, most structures of galaxies form early and have time for their substructures to "mix" (by mergers) or "melt" (by tidal disruption), whereas in an $\Omega = 1$ Universe, structures keep forming at all times, and one should see a higher fraction of clusters harboring important substructure. Recent studies by Ricshtone, Loeb & Turner (1992), Lacey & Cole (1993) and Kauffmann & White (1993) thus yield $\Omega \geq 0.5$. However a recent dynamical analysis indicates that substructures in clusters survive longer than was previously assumed, thus allowing lower values of $\Omega$ (Gonzalez-Casado, Mamon & Salvador-Solé 1993).

If clusters of galaxies are fair tracers of the distribution of baryons (gas and stars) in the Universe, then $\Omega = \Omega_b/f_b$ (White 1992), where $\Omega_b \simeq 0.06\,h_{50}^{-2}$ is the density of baryons (normalized to the density to close the Universe) obtained from standard big-bang nucleosynthesis (Steigman 1989), and $f_b$ is the fraction of baryons. X-ray analyses of clusters indicate baryonic fractions of $\simeq 25\%$ thus indicating $\Omega \leq 0.25\,h_{50}^{-2}$ (White 1992), while it should be mentioned that this baryonic fraction may be substantially lower in the inner regions of clusters (Hughes 1989; Durret *et al.* 1993).

The distribution of cluster temperatures serves as a powerful cosmological diagnostic, ruling out the Cold Dark Matter model. The favored models have either *mixed dark matter* (cold and hot) or an important *cosmological constant* (Bartlett & Silk 1993). The properties of clusters of galaxies seem to be correlated in a similar way as are the properties of elliptical galaxies and bulges of spirals, leading to a *fundamental plane* in the observed $(R, V, L)$ space, with nearly constant $M/L$ (Schaeffer *et al.* 1993).

Finally, whereas clusters are the preferred sites for elliptical galaxies, which are often thought to be the product of *mergers* (*e.g.*, Toomre 1977), it has been objected (Ostriker 1980) that the high velocity dispersions of clusters prohibit efficient merging, which requires slow encounters. A recent simple analytical analysis of the frequency of mergers in high velocity dispersion clusters (Mamon 1992) suggests that the observed fraction of ellipticals and its variation with environment can be explained by a small fraction of collisions that are slow enough to lead to merging.

## 3. Recent Developments on Groups

Much interest has arisen from the *ROSAT* detection of hot diffuse intergalactic gas in four groups (*ROSAT* has the ideal energy passband to detect such gas in groups). The first group with diffuse hot gas detected (NGC 2300) was reported to have a low baryonic fraction of 4% (Mulchaey *et al.* 1993), close to $\Omega_b$ (see §2), such that this group would be a fair tracer of an $\Omega = 1$ Universe. A reanalysis of this group shows that its baryonic fraction is over 20% (Henriksen & Mamon 1994), which extrapolates to $\Omega \leq 0.3$, if such groups are fair tracers of the baryon content of the Universe.

The biases in the estimates of the mass and crossing time of a group are function of its cosmo-dynamical state (expanding, collapsing, virialized ...). This implies a theoretical track, which can be mapped to the observed properties of groups, assuming a constant true $M/L$, providing a cut through a *fundamental surface* of groups (Mamon 1994). One derives $M/L = 440\,h(\rightarrow \Omega \simeq 0.3)$, four times larger than previous estimates from the same catalogs, because *most groups are relatively close to their cosmological turnaround*.

Although the interest in compact groups is growing perhaps exponential-ly, it is still unclear whether their 3D density is as high as it appears in projection. The galaxies in Hickson's (1982) compact groups often display unusual surface photometry (ellipticals: Mendes de Oliveira 1992) and internal kinematics (spirals: Rubin *et al.* 1991), so that in the 16 compact groups with complete observational analysis, 75% have at least 3 unusual galaxies (Mendes de Oliveira 1992), probably indicative of dynamical interactions.

If compact groups are *chance alignments* within larger loose groups of galaxies (Mamon 1986) one expects these to be *binary-rich* (Mamon 1992), Thus, the level of interactions in compact groups could be high without compact groups being dense in 3D. Indeed, the level of ongoing merging (Zepf 1993) and star formation (Moles *et al.* 1994) is much lower than for systems of interacting binaries, but higher than for isolated galaxies. Now assuming a mix of compact groups states, where most are binary-rich chance alignments, one expects at best 55% of groups with at least 3 galaxies in strong interaction (Mamon 1994), so there may be some (small) discrepancy with the high fraction in the 16 compact groups mentioned above.

An automated search for compact groups from a galaxy list obtained with the COSMOS scans of UKST plates has revealed differences with Hickson's compact groups (Iovino *et al.* 1993): The distribution of the magnitude difference between first and second ranked member is more skewed to high values in the automated catalog, suggesting that Hickson, who selected compact groups by eye on the POSS plates, was biased towards groups with 2 nearly equally bright dominant galaxies.

The fundamental surface analysis of groups provides clues to the nature of compact groups (Mamon 1994): from low to high velocity dispersion, compact groups are 1) chance alignments within collapsing loose groups, 2) loose groups near full collapse, 3) loose groups that have had time to collapse, virialize, mix their galaxy halos, and dissipate the orbital energy of their galaxies by dynamical friction against the merged common halo.

## 4. Applications of Near-IR Surveys

### 4.1 Immediate applications from a large-scale 2D survey

The work on groups and clusters of galaxies in the Near-IR (hereafter NIR) will not be just be a repeat of similar work performed in the optical bands. Indeed, the $K$ band has the advantage of being the optimum waveband to be at long enough wavelength to be relatively unaffected by the effects of obscuration from interstellar dust from our Galaxy or from other galaxies, and short enough to be unaffected by thermal emission from such dust. Also, the NIR is thought to be a better tracer of the stellar component in galaxies, in comparison with both optical and far-IR wavebands, which are more biased by star formation.

Therefore, DENIS and 2MASS near-IR surveys will provide unique oppor-tunities to probe at new questions, thanks to their two important attributes:

1) DENIS and 2MASS will provide the first large *near-IR selected* galaxy catalogs.

2) DENIS and 2MASS will be the first multi-color digital sky surveys (tied with the Sloan Digital Sky Survey, see Kent, in these proceedings).

The applications to groups and clusters are then:

1a) The first *automated near-IR selected* catalogs of clusters, loose groups, and compact groups. There are no such catalogs at present, except for compact groups (Iovino *et al.* 1993, see §3). Both APM and COSMOS teams plan to publish cluster catalogs, which they have used to study the large-scale cluster-cluster correlation function. The IRAS galaxy catalog has not been used to build catalogs of groups or clusters, because IRAS misses the elliptical galaxies that dominate the cores of rich clusters and are a significant component in groups. The NIR group and cluster calalogs should be free from the biases towards star-forming galaxies seen in other catalogs.

1b) The discovery of *groups and clusters at low galactic latitude*, for example in the direction of the Great Attractor, which is a mass concentration derived from large-scale velocity fields, but not well seen because of heavy dust obscuration at its low galactic latitude.

2a) Comparison of the properties of groups and clusters *versus waveband* ($I$, $J$, $H$, $K$). For example, will the properties of groups built from the $I$ band yield a different distribution of group collapse times than the properties of groups built from the $K$ band? If yes, hopefully this will not be a product of selection effects, but a trace of the underlying astrophysics.

2b) *Color segregation*. The first systematic study of the correlation of color with environment (even in 2D) will provide clues to the morphological evolution of galaxies. One will be constrained to high galactic latitudes to avoid differential extinction (reddening) causing spurious color segregation, and for similar reasons one may require avoiding the near edge-on spirals.

### 4.2  Followup studies

Of course, followups on DENIS and 2MASS will provide much greater insight into the astrophysics of groups and clusters. On one hand, a deeper NIR survey of some portion of the sky will allow the inclusion of dwarf galaxies in the group and cluster catalogs. On the other hand, a spectroscopic followup, which hopefully will eventually cover the full sky, will provide 3D information on groups and clusters, and thus eliminate a major fraction of the interlopers.

## 5.  DENIS Simulated View of Coma-Like Clusters

To illustrate the appearance of rich clusters, I present simulated images of the Coma cluster with the DENIS characteristics. The galaxy data has been obtained by scanning the LEDA galaxy database in a square of $2 \times 2°$, with the requirement of existing $B$ magnitude, morphological type, scale length and axis ratio. If the position angle was lacking it was assigned randomly. Each galaxy was divided into a bulge and a disk component, according to its morphological type. The blue magnitudes were converted to $J$ and $K$ magnitudes using $B - J = 1.9$ for disks and $3.2$ for bulges and $B - K = 2.8$ and $4.2$, respectively. The LEDA data is complete out to $B \simeq 15.5$ and the faintest galaxy is at $B = 17.4$.

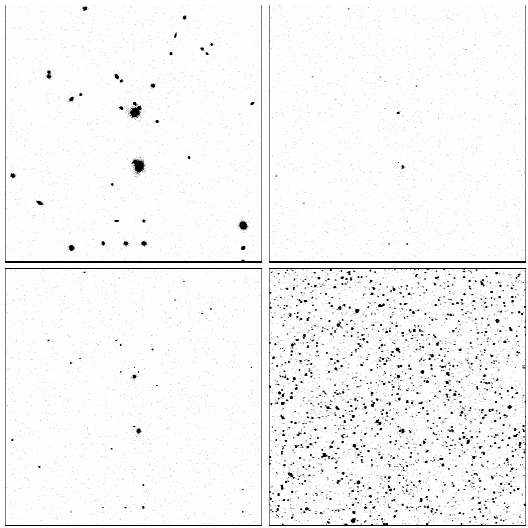

Fig. 1. Simulated images of the Coma cluster at 4 times its true distance, on a 1m telescope: $B(900\,\mathrm{s})$ (*upper left*), $K'$ (*upper right*), $J$ (*lower plots*); the latter three are in DENIS-like microscanned $9 \times 1\,\mathrm{s}$ integrations, the lower right plot is for a simulated star field analagous to that at $(\ell, b) = (0°, 5°)$, while the first three have no superposed stars.

Figure 1 shows 4 examples of this simulated Coma cluster, placed at 4 times its true distance from us (*i.e.*, $z \simeq 0.09$). The upper left figure is a $B$ band simulation with the same 1m telescope but with a 15 minute integration. The upper right and lower left plots are DENIS simulation in the $K'$ and $J$ bands, respectively. The lower right plot is the same $J$ band simulation but with a random star field of the same density as expected at $\ell = 0°, b = 5°$. For a direct comparison, all four images are grey-scaled from $2\sigma$ to $5\sigma$ above the theoretical sky background. The brightest galaxy has magnitudes $B = 15.5$, $J = 12.3$, and $K = 11.3$.

The $J$ image without a star field shows considerably more details than the $K'$ equivalent. The contamination from stars at low galactic latitudes appears sufficient to prevent an eyeball determination of cluster members.

## References

Bartlett, J.G. and Silk, J.: 1993, *Astrophysical Journal (Letters)* **407**, L45
Durret, F., Gerbal, D., Lachièze-Rey, M., Lima-Neto, G., and Sadat, R.: 1993, *Astronomy and Astrophysics* (submitted)
González-Casado, G., Mamon G.A., and Salvador-Solé E.: 1993, in preparation
Henriksen, M.J. and Mamon G.A.: 1994, *Astrophysical Journal (Letters)* (in press)
Hickson, P.: 1982, *Astrophysical Journal* **255**, 382
Hughes, J.P.: 1989, *Astrophysical Journal* **337**, 21
Iovino, A., Prandoni, I., MacGillavray, H.T., Hickson, P., and Palumbo, G.: 1993, in G. Chincarini, A. Iovino, T. Maccacaro and D. Maccagni (eds.), *Observational Cosmology*, A.S.P.: San Francisco, p. 276
Kauffmann, G. and White, S.D.M.: 1993, *Monthly Notices of the RAS* **261**, 921
Lacey, C.G. and Cole, S.: 1993, *Monthly Notices of the RAS* **262**, 627
Mamon, G.A.: 1986, *Astrophysical Journal* **307**, 426
Mamon, G.A.: 1992, *Astrophysical Journal (Letters)* **401**, L3
Mamon, G.A.: 1994, in F. Combes & E. Athanassoula (eds.), *Gravitational Dynamics and the N-Body Problem*, Meudon: Obs. de Paris, in press
Mendes de Oliveira, C.: 1992, Ph.D. thesis, Univ. of British Columbia
Moles, M. *et al.*: 1994, *Astronomy and Astrophysics* (in press)
Mulchaey, J.S., Davis, D.S., Mushotzky, R.F., and Burstein, D.: 1993, *Astrophysical Journal (Letters)* **404**, L9
Ostriker, J.P.: 1980, *Comm. Ap.* **8**, 177
Richstone, D., Loeb, A., and Turner, E.L.: 1992, *Astrophysical Journal* **393**, 477
Salvador-Solé, E., Sanromà, M., and González-Casado, G.: 1993, *Astrophysical Journal* **402**, 398
Salvador-Solé, E., González-Casado, G. and Solanes, J.M.: 1993, *Astrophysical Journal* **410**, 1
Schaeffer, R., Maurogordato, S., Cappi, A., and Bernadeau, F.: 1993, *Monthly Notices of the RAS (Letters)* **263**, L21
Steigman, G.: 1989, C. Jake Waddington (ed.), in *Cosmic Abundances of Matter*, AIP: Minneapolis, 310
Toomre, A.: 1977, in B.M. Tinsley and R.B. Larson (eds.), *The Evolution of Galaxies and Stellar Populations*, Yale Univ. Obs.: New Haven, 401
Walke, D.G. and Mamon, G.A.: 1989, *Astronomy and Astrophysics* **295**, 291
White, S.D.M.: 1992, in A.C. Fabian (ed.) *Clusters and Superclusters of Galaxies*, Kluwer: Dordrecht, 17
White, S.D.M., Briel, U.G. and Henry, J.P.: 1993, *Monthly Notices of the RAS (Letters)* **261**, L8
Zepf, S.E.: 1993, *Astrophysical Journal* **407**, 448